\documentstyle[12pt]{article}
\begin{document}
  \def\theequation{\arabic{section}.\arabic{equation}} 
  \newcommand{\be}{\begin{equation}}
  \newcommand{\ee}{\end{equation}}
  \begin{titlepage}
  \title{The influence of the cosmological expansion \\ on local systems}
  \author{F. I. Cooperstock$^1$, V. Faraoni$^2$  and D. N. Vollick$^1$\\ 
          \\{\small \it $^1$Department of Physics and Astronomy,
University 
  of Victoria} \\
  {\small \it P.O. Box 3055, Victoria, B.C. Canada V8W 3P6 }\\\\
{\small \it Inter--University Centre for Astronomy and Astrophysics}\\
{\small \it Post Bag 4, Ganeshkhind, Pune 411 007, India}}
  \date{
  } 
  \maketitle
  \thispagestyle{empty}
  \end{titlepage}  \clearpage
  \begin{center} {\bf ABSTRACT} \end{center}
  
  Following renewed interest, the problem of whether the cosmological 
  expansion affects
  the dynamics of local systems is reconsidered. The cosmological correction to
  the equations of motion in the locally inertial Fermi normal frame (the 
  relevant frame for
  astronomical observations) is computed. The evolution equations for the
  cosmological perturbation of the two--body problem are solved in this frame.
  The effect on the orbit is insignificant as are the effects on the galactic
  and galactic--cluster scales.
  \vspace*{1truecm}
  To appear in the Astrophysical Journal

  \vspace*{1truecm}
  \vspace*{0.5truecm}  \begin{center}  Keywords: Cosmology: 
  theory~--~relativity~--~celestial mechanics, stellar dynamics. \end{center}
  \clearpage
  
  \section{Introduction}
  
  A recurring issue in cosmology concerns the nature and extent of the
  cosmological expansion. If expansion were to occur in proportion and in
  every minutia of detail to every element of the universe, then 
  every clock and every measuring apparatus of distance would be 
  altered in proportion. If in addition, the laws of physics were to 
  remain unaltered in the process, the very concept of expansion would lose
  its meaning as it would be intrinsically unobservable. 
  This is not what is
  being contemplated for our universe since we observe a systematic redshift of
  distant galaxies and hence we
  are able to deduce that there is an expansion in progress, at least on the
  cosmological scale. While the effect is actually registered in the small 
  distances of a 
  wavelength of light, this is simply an imprint of the expansion at the 
  largest (Hubble) scale. 
     
  Recently (Anderson 1995; Bonnor 1996) there has been a revival of
  interest in the question as to whether the cosmological expansion also 
  proceeds at smaller scales. There is a tendency to reject such an 
  extrapolation by confusing it with the intrinsically unobservable "expansion"
  (let us refer to this as "pseudo-expansion") described above.
  By contrast, the metric of Friedman--Robertson--Walker (FRW) in
  general relativity is intrinsically dynamic with the increase (decrease)
  of proper distances correlated with red--shift (blue--shift). It does so
on any scale provided the light travel time is much longer than the wave
period. Thus, the cosmological metric alone does not dictate a scale for
expansion and in principle, it could be present at the smallest practical
scale as real--  as opposed to pseudo--expansion, and observable in principle.
  
However, it is reasonable to pose the question as to
   whether there is a cut--off at
which
  systems below this scale do not partake of the expansion. It would
appear that
  one would be hard put to justify a particular scale for the onset of
expansion. 
         Thus, in this debate, we are in agreement with Anderson (1995) 
  that it is most
  reasonable to assume that the expansion does indeed proceed at all scales.
  However, there is a certain ironical quality attached to the debate in the
  sense that even if the expansion does actually occur at all scales, we will
  show that the effects of the cosmological expansion on smaller spatial and
  temporal scales would be undetectable in general in the foreseeable 
  future and hence one could just as comfortably hold the view that the 
  expansion occurs strictly on the cosmological scale.

       The question of whether the expansion of the universe affects local
  systems like clusters of galaxies or planetary systems was first raised
  many years ago and has received continued scrutiny (McVittie 1933; 
  J\"{a}rnefelt
  1940, 1942; Pachner 1963; Dicke \& Peebles 1964; Callan {\em et al.} 1965; 
  Irvine 1965; Noerdlinger
  \& Petrosian 1971) with the most recent consideration by Anderson (1995) 
  who extends the question to the stellar scale and even below this. The 
  recurrent
  attention paid to this issue indicates that to this point a definitive answer
  is still lacking. However, it is our sense that the prevalent perception 
  is that the physics of systems which are small compared to the radius of
  curvature of the cosmological background is essentially unaffected by the
  expansion of the universe.
  
  In the presence of spherical symmetry, the 
  analysis of a spherical cavity embedded in an
  FRW universe is well known: as a consequence of
  Birkhoff's theorem, the metric inside the spherical cavity is the Minkowski
  one, and the physics is the same as in flat space (Einstein \& Straus 1945;
  Sch\"{u}cking 1954; Dicke \& Peebles 1964;
  Callan {\em et al.} 1965; Bonnor 1996). However, when spherical 
  symmetry is absent, a
  satisfactory {\em quantitative} answer is missing in the literature, and
  certain statements about small systems being sheltered from the cosmological
  expansion recur (see e.g. Misner {\em et al.} 1973
  and the discussion in Anderson 1995). Noerdlinger \& Petrosian
  provide a quantitative treatment, but it is limited to the particular problem 
  of the collapse of galaxy clusters. Anderson's (1995) paper employs the
  Einstein--Infeld--Hoffmann method to derive the cosmological corrections to 
  the
  equations of motion of a system of particles subject to external forces. 
  When the dynamics of a single particle are considered, the 
  correction to the particle's acceleration is found to be proportional to 
  the velocity of the particle.
  However, it is 
  not clear how to relate the coordinates used by  Anderson
to the coordinates used by observers making astronomical observations.
  This is an important 
  issue because the computation 
  does not provide a coordinate--invariant
  quantity, but rather a correction to the 3--dimensional
  equations of motion, that are dependent upon the chosen coordinate
  system.
  Bonnor (1996) studies a distribution of pressureless charged dust in
  equilibrium between electrical repulsion and gravitational attraction, and
  concludes that it participates in the universal expansion.

  A qualitative answer to the problem of whether local systems are
affected by the
  expansion of the universe is easily provided if one considers the
  equivalence principle and its geometric formulation. 
  Although the cosmological expansion is described by the time--dependent scale
  factor in the FRW metric, and we believe affects lengths at all scales, 
  the
  curved spacetime manifold can be locally approximated by its (flat) tangent
  space at every spacetime point $p$. This approximation is valid only in a 
  neighborhood $U(p)$ of the point $p$ considered; the error involved in the
  approximation increases with the size of the neighborhood $U(p)$, and the
  approximation breaks down completely when the size of $U(p)$ becomes 
  comparable to the radius of curvature  of spacetime (the Hubble radius 
  in the case of an 
  FRW spacetime). From the physical point of view, the
  tangent space at $p$ describes the spacetime seen by a freely falling
observer
  in the so--called locally inertial frame (hereafter called ``LIF''-- see
Landau
  \& Lifshitz 1989). This frame is the one in which astronomical
  observations are carried out. Thus, the effect of the cosmological
expansion is
  seen to be negligible locally and grows in significance with distance,
reaching
  full import on the cosmological scale. This conclusion is qualitative,
  and is certainly well--known to most relativists but, to the best of the
  authors' knowledge, has yet to be well--formulated quantitatively. In
earlier
  treatments, the coordinate systems adopted do not correspond to those
used by a
  physical observer. 
  
  The purpose of the present paper is to provide a
  clear quantitative answer to the problem. The motion of a particle
  subject to external forces in the (approximate) LIF using Fermi normal
  coordinates is analyzed. It is the locally inertial frame based on a
geodesic observer and it continues to be locally inertial following the
observer in time. 
This is
  the frame in which
  astronomical observations are performed, and we compute the corrections
to the
  dynamics due to cosmology. In this paper, we assume that homogeneous
isotropic expansion is actually universal and we analyze the consequences
of this assumption. 
  
  The plan of the paper is as follows: in Sec.~2 the equation of
  timelike geodesics in the LIF in an Einstein--de Sitter universe is
investigated
  and the cosmological perturbations to the 3--dimensional equations of
motion
  in the LIF are derived. In Sec.~3
  the orders of magnitude of the effects in realistic astrophysical
  systems are estimated, and it is demonstrated that they are very small and 
  unobservable with present and foreseeable technology. In Sec.~4 the two--body 
  problem in the LIF is studied in detail using the correction to the equations
  of motion computed in Sec.~3, thus providing a solution to the evolution
  equations for the perturbations. It is shown that cumulative effects of the
  cosmological expansion on the present orbital radius of the earth and its
  orbital motion are essentially negligible. Section~5 contains 
  a discussion and the conclusions.
  
\section{Equations of motion in the LIF}
In this section we find the equations of motion for a particle in the LIF 
using the geodesic deviation equation. We refer the reader to the Appendix
for the details of the calculation. 

 The metric in FRW coordinates for an Einstein -- de Sitter universe is
given by
\begin{equation}
ds^2=-dt^2+a(t)^2 \left[ dx^2+dy^2+dz^2 \right] \; ,
\label{metric}
\ee
where $a(t)$ is the scale factor. Consider an observer whose world line
is the geodesic $t=\tau$ and $\vec{r}=0$.
In Fermi normal 
coordinates, the metric on the geodesic is $g_{\mu\nu}=
\eta_{\mu\nu}$ and a parallely propagated orthonormal tetrad is given by
\be
\hat{e}_t^F=(1,0,0,0)\;\;\;\;\;\;\; \hat{e}_x^F=(0,1,0,0)
\ee
\be
\hat{e}_y^F=(0,0,1,0)\;\;\;\;\;\;\; \hat{e}_z^F=(0,0,0,1) \; .
\ee
The FRW basis vectors $\vec{e}^{FRW}_{\nu}$ are related to the $\hat{e}^F
_{\mu}$ via
\be
\hat{e}^F_{\mu}=\Lambda_{\mu}^{\;\;\nu}\vec{e}_{\nu}^{FRW}
\ee
where $\Lambda_{\mu}^{\nu}=$diagonal(1,$a^{-1},$ $a^{-1},$ $a^{-1}$). The
Riemann
tensor in Fermi normal coordinates $\{ x_F^{\mu} \} $, along the geodesic,
is given by (Greek indices range from 0 to 3, Latin indices from 1 to 3)
\be
R^F_{\alpha\beta\mu\nu}=\Lambda_{\alpha}^{\;\;\sigma}\Lambda_{\beta}^
{\;\;\lambda}\Lambda_{\mu}^{\;\;\omega}\Lambda_{\nu}^{\;\;\kappa}R^{FRW}_{
\sigma\lambda\omega\kappa}
\ee
(where the superscript $F$ denotes quantities in Fermi normal coordinates)  
and the geodesic deviation equation
\be
\frac{d^2x^k}{d\tau^2}+\Gamma^k_{\alpha\beta}\frac{dx^{\alpha}}{d\tau}
\frac{dx^{\beta}}{d\tau}+{R^k}_{0l0}x^l=0
\ee
becomes
\be
\frac{d^2x^k_F}{d\tau^2}+{R^k}_{0l0}x^l_F=0
\ee
in Fermi normal coordinates, since $\Gamma^{(F) \mu}_{\alpha\beta}=0$.
Thus, to lowest order in $x^{\mu}$ and $dx^{\mu}/d\tau$ (the order to
which the geodesic deviation equation is valid), the equations of motion
in Fermi normal coordinates are (see the Appendix)
\be
\frac{d^2 x^k_F}{d\tau^2}-\left(\frac{\ddot{a}}{a}\right)x^k_F =0 \; ,
\label{eqofmot}
\end{equation}
where a overdot denotes differentiation with respect to the comoving time 
$t$.
  \section{Order of magnitude estimates}
  
\setcounter{equation}{0}
  In this section, the order of magnitude of the
  effect created by the cosmic expansion on the dynamics of local systems is
  estimated. Astronomical systems for which the velocities involved are 
  non--relativistic are considered. The 
  present value of the age of the universe is taken to be 
  $6.3\times 10^{17}$~seconds.

{\bf Acceleration on the scale of the Solar System}\\
  It is sufficient for our purposes to use the present value of the average
  size of the earth-sun system, i.e. the astronomical unit $r_0$ of $1.5
\times 10^{11}$~m and the present orbital frequency $\omega_0$ of 
$2\times 10^{-7}$ sec$^{-1}$. From eq. (\ref{radius}) of the following
section, the correction to
the acceleration for this distance and frequency due to the cosmological
expansion at the present matter--dominated ($a(t) \propto t^{2/3}$) epoch
is
\be
\delta \ddot{r}=-\frac{4r_0}{3t^4\omega_0^2}=-3.17\times 10^{-47}
{\mbox m}/\mbox{sec}^2   \; .
\ee
This is to be compared to the predominant gravitational acceleration of
the earth towards the sun
\be
g=\frac{GM_{\odot}}{r_0^2}=6\times 10^{-3} {\mbox m}/\mbox{sec}^2
\ee
which completely overwhelms the effect of the cosmological expansion by 44
orders of magnitude. This is in qualitative agreement with a cruder order
of magnitude estimate in Lightman {\em et al.} (1975).

  {\bf Acceleration on the galactic scale} \\
  As an example, consider our galaxy, a spiral in which the sun is located at
  $r_0=8.5 \pm 1$~Kpc from the center and has orbital velocity
  $v_0=220 \pm 15$~Km~s$^{-1}$ (Binney \& Tremaine 1987). 
  
  Thus, the gravitational acceleration of the sun towards the center of the
  galaxy $g=v_0^2/r_0 \simeq 1.9\times 10^{-10}$ m/sec$^2$. 
Since the orbital period is
  $\simeq 2\times 10^8$ yrs, the angular velocity $\omega_0$ 
is of the order $10^{-15}$
  sec$^{-1}$. Thus, the correction of the acceleration of the sun due to the
  cosmological expansion
\be
\ddot{r}=-\frac{4r_0}{3t^4\omega_0^2}
\ee
is of the order $10^{-21}$ m/sec$^2$ at the present epoch which is 11
orders of magnitude smaller than the galactic $g$.

  {\bf Acceleration on the galactic cluster scale}\\
  Assuming the core radius of a galaxy cluster ($r_0\sim
  250$~Kpc) and the line of sight velocity dispersion $\sigma \simeq
  800$~Km~s$^{-1}$ (Binney \& Tremaine 1987) and assuming that the
  galaxies at the edge of the core of the cluster are in orbit around the 
  centre of
  the core with velocity $\sigma$, they are subject to the gravitational 
  acceleration $g=(v_0)^2/r_0\simeq 8 \times 10^{-11}$~m~s$^{-2}$. This is
  to be compared with the correction due to the cosmological expansion \be
\ddot{r}=-\frac{4r_0^3}{3t^4v^2}=-5.6\times 10^{-18} {\mbox
m}/\mbox{sec}^2 \; .  \ee 
While this is 7 orders of magnitude smaller than the
galactic cluster $g$ and thus of considerably greater relative
significance than was found for the galactic and the solar system scales,
it is still nevertheless essentially ignorable.

\section{Cosmological corrections to the two--body problem in the LIF}
  
  The effects of the expansion of the universe on the dynamics of local systems
  are exemplified by the corrections induced in the two--body problem. 
  The two--body problem in a cosmological background has been
  analyzed  in previous papers (McVittie 1933; Dicke \& Peebles 1964; 
  Noerdlinger \& Petrosian
  1971; Anderson 1995) with differing results. McVittie (1933) reached the
  conclusion that the orbital radius stays constant for an observer using
  coordinates fixed in the solar system\footnote{See Ferraris {\em et al.}
(1996) 
  for a modern criticism of McVittie's coordinates in astrophysical
  applications.}. Dicke \& Peebles (1964) used a
  conformal technique to show that the coordinate radius of the orbit
decreases as the
  inverse of the scale factor and that the proper radius stays constant. 
  Noerdlinger \& Petrosian (1971) considered the
  two--body problem inside a cluster and found that, in a
  dust--dominated FRW universe, the time derivative of the average orbital
  radius obeys the equation \setcounter{equation}{0}
\be \label{41}
  \dot{\langle r \rangle}=\frac{3\epsilon}{1+4\epsilon} \, H\langle r
  \rangle \; ,
\ee
where $\epsilon$ is the ratio of average energy densities in the cluster 
and in the rest of the universe (compare eq.~(8) of Noerdlinger \&
Petrosian 1971).  According to this result, the orbital radius increases
and the effect is proportional to $\left| H_0 x^i\right|$. The more recent
result of Anderson (1995) agrees with that of Dicke \& Peebles (1964), but
in different coordinates. A comparison of all these results is rendered
difficult by the
  different coordinate systems adopted in the different studies. Moreover, 
  no treatment of the problem was given in the LIF, which is the 
  frame of reference relevant for astronomical observations performed by a 
  freely falling observer. In fact, the 3--dimensional
  equations of motion of a particle are not coordinate--invariant
  and, like the equations of motion themselves, the
  correction due to the cosmic expansion is dependent upon the frame employed. 
  In this section, we apply the
  results obtained in Sec.~2 to compute the perturbations of the two--body
  problem in the LIF in an expanding, matter--dominated Einstein--de
Sitter
universe. For simplicity, we restrict ourselves to the case of circular
orbits, in which the equation of motion for the two--body problem takes
the form
\be
\frac{d^2\vec{r}}{dt^2}-\frac{\ddot{a}}{a}\, \vec{r}=-\frac{GM}{r^2} \,
\underline{e_r}                               \label{grav}
\ee
  where $M$ is the mass of the central object. In this section we only use
  quantities defined in the LIF, and we drop the subscripts. Cylindrical coordinates 
  $\left( r, \theta, z \right)$
  are used, with associated unit vectors
  $\underline{e_r}$, $\underline{e_{\theta}}$ and $\underline{e_z}$. Since
the perturbation of the central force is also central, the
  motion is again confined to the unperturbed orbital plane. 
We consider the perturbation of the orbital coordinates $r$, $\theta$
given by 
\be \label{46}
  r(t)=r_0+\delta r(t) \; ,
\ee
\be \label{47}
\theta (t)=\omega_0 t+ \delta \theta (t) \; .
\ee 
Substitution into eq.~(\ref{grav}) yields
\be
r_0\delta\ddot{\theta}+2\delta\dot{r} \omega_0 =0 \; ,
\label{4.5}
\ee
\be
\delta\dot{\theta}=-\frac{2\omega_0}{r_0} \, \delta r \; ,
\label{4.6}
\ee
\be
\delta\ddot{r}-3\omega_0^2\delta r-2\omega_0 r_0\delta\dot{\theta}-
\frac{\ddot{a}}{a} \, r_0=0
\label{4.7}
\ee
where
\be
\frac{\ddot{a}}{a}=-\frac{2}{9t^2}
\label{4.8}
\ee
in a matter--dominated universe. Combining eqs. (\ref{4.6})--(\ref{4.8})
yields 
\be
\delta \ddot{r}+\omega_0^2\delta r+\frac{2}{9t^2}r_0=0.
\label{4.9}
\ee
It is easy to show that $\delta\ddot{r}$ is negligible relative to the
other two terms in eq.~(\ref{4.9}) when $t$ is of the order of the age
of the universe and 
  $\omega_0^{-1}$ is of the order of a year. Thus, we find that 
\be      \label{radius}
 r(t)\simeq r_0\left[ 1-\frac{2}{9t^2\omega_0^2}\right] \; .
 \label{4.10}
\ee
For the earth--sun system, we take $r_0 \simeq 1.5\times 10^{11}$~m  
(although strictly speaking, it is actually the value of $r_0$ as $t$
approaches infinity).
The value of $t$ is taken as the age of the universe, approximately
$2\times 10^{10}$ years or $6.3\times 10^{17}$ sec. The angular frequency
$\omega_0$ is taken to be for the earth--year, approximately $2\times
10^{-7}$ sec$^{-1}$.

From eqs. (\ref{4.6}) and (\ref{4.10}),
\be
\delta\dot{\theta}=\frac{4}{9t^2\omega_0}   \; .
\ee
Thus:
  
  1)~The angular velocity {\em decreases} with time: as $t$ approaches
infinity, $\omega$ approaches $\omega_0$ so strictly speaking, $\omega_0$
is actually the terminal angular velocity.
  
  2)~The orbit size {\em grows} with time: as $t$ approaches infinity, $
r$ approaches $r_0$.
  
  Now consider the fractional rate of change of frequency:
\be
\left[\frac{\dot{\omega}}{\omega}\right]_{present}=-8.9\times
10^{-41}\mbox{sec}^{-1} =-2.8\times 10^{-33} \mbox{yr}^{-1} \; .
\ee
  This can be compared to the observed rate of variation of the orbital  period
  of the moon about the earth, $2.22\pm 0.35 
\times 10^{-11}$ yr$^{-1}$, which is larger
  by approximately 22 orders of magnitude. 
  The cosmological effect is not significantly different at the birth of the 
  solar system. For $t$, we would use the present time minus the age of
the solar system which is still of order $10^{17}$ sec. Thus the rate at birth
  of the solar system was not significantly different.

For the fractional change in radius of the orbit, we use eq.~(4.6) to find
\be
\frac{\delta\dot{r}}{r_0}=-\frac{\delta\ddot{\theta}}{2\omega_0} \; ,
\ee
  which gives the same kind of insignificant rate of radius growth with
  the expansion of the universe. Over the life span of the solar system, 
of order $10^{17}$~sec, the fractional change in radius was a mere 
$10^{-24}$.
  
  \section{Discussion and conclusions}
  
  The effect of the cosmic expansion on the dynamics of local spherically
  symmetric systems is well--known (Einstein \& Straus 1945; Dicke \&
Peebles
  1964; Callan {\em et al.} 1965; Bonnor 1996). In the non--spherical
case, it is
  generally recognized that the expansion of the universe does not have
  observable effects on local physics, but few discussions of this problem
in the
  literature have gone beyond qualitative statements. A serious problem
  is that these
  studies were carried out in coordinate systems that are not easily
comparable
  with the frames used for astronomical observations and thus obscure the
physical
  meaning of the computations. Moreover, different treatments lead to
apparently
  conflicting results, as in the case of the two--body problem. This is
the
  reason why the computations of Secs.~2 and 4, performed in the LIF, 
  are particularly relevant to the problem. While it is reasonable to
assume that
  the time dependence of the scale factor  in the FRW metric
  (\ref{metric}) affects lengths at all scales in principle (see the discussion 
  in Anderson 1995; Bonnor 1996), the magnitude of the effect in the LIF is
  the physically relevant one, and its computation constitutes the essential
  aspect of this work.

  The
  computation of the cosmological correction to the local equations of
motion
  performed in Sec.~2 allows one to estimate numerically the magnitude of
the
  correction to the acceleration of a particle subject to external forces.
The
  numerical estimates obtained in Sec.~3 suggest that the correction is
extremely
  small and
  unobservable for galaxy clusters, galaxies and the solar system, and
  negligible for smaller systems such as stars and even more so for
molecules and
  atoms (cf. Anderson 1995). When the cosmological correction
  to the local equations of
  motion is applied to the Newtonian two--body problem, the evolution
equations
  for the perturbation of the orbit can be solved. It is found that the
  cumulative effect of cosmological expansion on the radius and angular
motion of
  the sun--earth system is also negligible. The cosmic expansion plays an
  increasingly important role for systems whose sizes and
  lifetimes become increasingly comparable to the Hubble radius and to
Hubble
  times respectively. In this case, the
  approximation used in this paper becomes invalid. It is
  well--known that the cosmological expansion must be taken into account,
for
  example, in the fluid dynamical treatment of the formation of structures
in
  the universe (Weinberg 1972). As a conclusion, it is reasonable to
assume that
  the expansion of the universe
  affects all scales, but the magnitude of the effect is 
  essentially negligible for local systems, even at the scale of galactic 
  clusters.

  \section*{Acknowledgments}
  
  We are grateful to Prof. W.B. Bonnor and Prof. E.L. Wright 
  for helpful discussions. 
  This research was supported, in part, by a grant from the Natural Sciences 
  and Engineering Research Council of Canada. 
  \clearpage
  
\section*{Appendix}

In this appendix we find the transformations from FRW coordinates to
Fermi normal coordinates. We also find the metric to order $|\vec{x}|^2$
and the equations of motion to lowest order.
  
Consider an observer whose world line is the geodesic
$r=0$. To find the Fermi normal coordinates of a 
point $P=(t_{FRW},\vec{x}_{FRW})$ we find the unique spacelike geodesic
which goes through the point P and intersects $r=0$ orthogonally. For a
sufficiently small region about $r=0$, such a unique geodesic is
guaranteed
to exist. Let Q be the point of intersection between $r=0$ and this
geodesic and let the geodesic parameter $\tau$ be zero at Q.
The initial velocity vector $T^{\mu}=dx^{\mu}/d\tau|_{\tau=0}$
is chosen so that the geodesic reaches P at $\tau=1$. The Fermi normal time
$t_F$ is taken to be the proper time from the 
initial cosmological singularity to the point Q
along the geodesic $r=0$. The Fermi normal spatial coordinates are given by the
projection of $T^{\mu}$ onto the orthonormal triad $e^{\mu}_{(a)}$, where
$e^{\mu}_{(1)}, e^{\mu}_{(2)},$ and $e^{\mu}_{(3)}$ point in the $x$, $y$,
and $z$ directions respectively.

\def\theequation{A.\arabic{equation}}
\setcounter{equation}{0}
   
The geodesic equation has solutions of the form
\begin{equation}
x^k=x^k(\tau,c_m)\;\;\;\;\;\;\;\;\;\;\;\;\;\;\; t=t(\tau,c_m)
\end{equation}
where $\{c_m\}$ is a set of eight constants. From the above discussion we
have the following conditions
\begin{equation}
x^k(0,c_m)=0\;\;\;\;\;\;\;\;\;\;\;\;\; t_F=t(0,c_m) \; ,
\end{equation}
\begin{equation}
\frac{\partial t}{\partial\tau}(0,c_m)=0\;\;\;\;\;\;\;\;\; x_F^k=
a(t_F)\frac{\partial x^k}{\partial\tau}(0,c_m) \; .
\end{equation}
This set of equations allows us to solve for $c_m=c_m(x_F^k,t_F)$.
Substituting this into
\begin{equation}
x^k_{FRW}=x^k(1,c_m)\;\;\;\;\;\;\;\;\; t_{FRW}=t(1,c_m) 
\end{equation}
gives the required transformations
\begin{equation}
x^k_{FRW}=x^k_{FRW}(x_F^m,t_F)\;\;\;\;\;\;\;\;\;
t_{FRW}=t_{FRW}(x^m_F,t_F) \; .
\end{equation}
   
The geodesic equations are
\begin{equation}
\frac{d^2t}{d\tau^2}+a\dot{a}\left[\left(\frac{dx}{d\tau}\right)^2+
\left(\frac{dy}{d\tau}\right)^2+
\left(\frac{dz}{d\tau}\right)^2\right]=0 \; ,
\label{ap1}
\end{equation}
and
\begin{equation}
\frac{d^2x^k}{d\tau^2}+2\, 
\frac{\dot{a}}{a} \, \frac{dt}{d\tau} \, \frac{dx^k}{d\tau}=0 \; .
\label{ap2}
\end{equation}
From eq.~(\ref{ap2}) we have
\begin{equation}
\frac{dx^k}{d\tau}=\frac{C_1^k}{a^2} \; , 
\end{equation}
where the $C_1^k$ are constants. From equation (\ref{ap1}) we have
\begin{equation}
\frac{dt}{d\tau}=\sqrt{C_2+\frac{|\vec{C_1}|^2}{a^2}} \; , 
\end{equation}
where $C_2$ is a constant.
  
We now specialize to FRW spacetimes with 
\begin{equation} 
a(t)=(\alpha t)^n 
\end{equation} 
where $\alpha$ and $n$ are constants. The above
differential equations have the power series solutions 
\begin{equation}
t(\tau)=t_0+\sqrt{C_2+\frac{|\vec{C}_1|^2}{(\alpha t_0)^{2n}}}\;\tau
-\frac{n|\vec{C}_1|^2}{2 ( \alpha
t_0)^{2n}t_0}\;\tau^2+\frac{(2n+1)n|\vec{C}_1|^2} {6 (\alpha
t_0)^{2n}t_0^2}\sqrt{C_2+\frac{|\vec{C}_1|^2}{(\alpha t_0)^{2n}}}\;\tau^3
+O(\tau^4) \; , 
\end{equation} 
\begin{eqnarray} x^k(\tau) & = &
x^k_0+\frac{C_1^k}{(\alpha t_0)^{2n}}\;\tau -\frac{n C_1^k}{(\alpha t_0)^
{2n}t_0}\sqrt{C_2+\frac{|\vec{C}_1|^2}{(\alpha t_0)^{2n}}}\;\tau^2
\nonumber \\ & & +\frac{
n C_1^k \left[(2n+1)C_2+(3n+1)\frac{|\vec{C}_1|^2}{(\alpha
t_0)^{2n}}\right]} {3(\alpha t_0)^{2n}t_0^2} \;\tau^3+O(\tau^4) \; ,
\end{eqnarray} 
where $t_0$ and $x_0^k$ are constants. Now $t_F=t(\tau=0)$
gives $t_F=t_0$, $x^k(\tau=0)=0$ gives $x_0^k=0$,
$x_F^k=a(t_0)dx^k/d\tau|_{\tau=0}$ gives $C_1^k=(\alpha t_F)^nx_F^k$, and
$dt/d\tau|_{\tau=0}=0$ gives $C_2=-|\vec{x}_F|^2$.  Thus, using
$t_{FRW}=t(\tau=1)$ and $x^k_{FRW}=x^k(\tau=1)$, we have 
\begin{equation}
t_{FRW}=t_F-\frac{n|\vec{x}_F|^2}{2t_F}+{\mbox O} (|\vec{x}_F|^4) 
\end{equation} 
and 
\begin{equation} 
x^k_{FRW}=\frac{x_F^k}{(\alpha
t_F)^n}   \left[ 1+\frac{n^2|\vec{x}_F|^2}{3 t_F^2}\right] +{\mbox
O}(|\vec{x}_F|^4) \; .  
\end{equation} 
Note that to lowest order
$x_F^k=a(t_{FRW})x^k_{FRW}$ and $t_F=t_{FRW}$, so that to lowest order
Fermi normal coordinates are just ``physical'' coordinates in FRW
spacetime. 
  
The spatial components of the geodesic equation, to lowest order in $x^k$
and $\dot{x}^k$, are \begin{equation}
\frac{d^2\vec{x}_F}{dt^2}-\frac{\ddot{a}}{a}\, \vec{x}_F=0 \end{equation}
which is identical to (\ref{eqofmot}). The metric in Fermi normal
coordinates is 
\begin{equation}
ds^2=-\left[1-\frac{n(n-1)|\vec{x}_F|^2}{t_F^2}\right]dt^2_F+
\left[\delta_{kl}\left(1-
\frac{n^2|\vec{x}_F|^2}{3t_F^2}\right)+\frac{n^2x_k^Fx_l^F}{3t^2_F}
\right]dx_F^kdx_F^l \; .  
\end{equation} 
This can be written, to lowest
order in $x_F$, as 
\begin{equation}
ds^2=-(1+R^F_{0l0m}x^l_Fx^m_F)dt^2_F-\left(\frac{4}{3}R^F_{0ljm}x_F^lx_F^m
\right)dt_F dx^j_F+\left[ \delta_{ij}-\frac{1}{3}R^F_{iljm}x^l_Fx^m_F
\right] dx^i_Fdx^j_F 
\end{equation} 
since, to lowest order, the nonzero
components of the Riemann tensor are 
\begin{equation}
R^F_{0x0x}=R^F_{0y0y}=R^F_{0z0z}=-\,
\frac{\ddot{a}}{a}=-\frac{n(n-1)}{t_F^2} 
\end{equation} 
and
\begin{equation}
R^F_{xyxy}=R^F_{xzxz}=R^F_{yzyz}=\left(\frac{\dot{a}}{a}\right)^2=\frac{n^2}
{t_F^2} \end{equation} (plus components related to these by symmetry).
This expression is identical to the metric in Fermi normal coordinates
given by Manasse and Misner (1963). 
  \clearpage
  
  \section*{References}
  
  \noindent Anderson, J. L. 1995, Phys. Rev. Lett. 75, 3602
  
  \noindent Binney, J. \& Tremaine, S. 1987, Galactic Dynamics (Princeton:
  Princeton University Press)
  
  \noindent  Bonnor, W. B. 1996, MNRAS 282, 1467
  
  \noindent Callan, C. {\em et al.} 1965, Am. J. Phys. 33, 105
  
  \noindent Dicke, R. H. \& Peebles, P. J. E. 1964, Phys. Rev. Lett 12, 435
  
  \noindent Einstein, A. \& Straus, E. G. 1945, Rev. Mod. Phys. 17, 120
  
  \noindent Ferraris, M. {\em et al.} 1996, Nuovo Cimento 111B, 1031
  
  \noindent Irvine, W. M. 1965, Ann. Phys. (NY) 32, 322
  
  \noindent J\"{a}rnefelt, G. 1940, Ann. Acad. Sci. Fenn. Ser. A, 55, Paper 3
  
  \noindent ----------------- 1942, Ann. Acad. Sci. Fenn. Ser. A, 1, Paper 12
  
  \noindent Landau L. D. \& Lifshitz E. M. 1989, The Classical Theory of Fields,
  fourth revised edition (Oxford: Pergamon Press)

  \noindent Lightman A. P., Press W. H., Price R. H. and Teukolsky S. A.
  1975, Problem Book in Relativity and Gravitation (Princeton: Princeton
  University Press)

  \noindent Manasse, F. K. and Misner, C. W. 1963, J. Math Phys. 4, 735
  
  \noindent McVittie, G. C. 1933, MNRAS 93, 325
  
  \noindent Misner, C. W. {\em et al.} 1973, Gravitation (San Francisco: Freeman)
  
  \noindent Noerdlinger, P. D. \& Petrosian, V. 1971, ApJ 168, 1
  
  \noindent  Pachner, J. 1963, Phys. Rev. 132, 1837
  
  \noindent Sch\"{u}cking, E. 1954, Z. Phys. 137, 595
  
  \noindent Van Flandern, T. C. 1975, MNRAS 170, 333
  
  \noindent Weinberg, S. 1972, Gravitation and Cosmology (New York: J. Wiley \&
  Sons)
  \end{document}